\documentclass{elsart}

\usepackage{epsfig}
\usepackage{times,mathptmx}

\begin{document}
\newpage
\begin{frontmatter}
\title{Temperature inversion in granular fluids under gravity}

\author[CECAM]{Rosa Ram\'{\i}rez}
\ead{roramire@cecam.fr}
\author[CECAM,UChile]{Rodrigo Soto}
\address[CECAM]{CECAM, ENS-Lyon, 46 All\'ee d'Italie, 69007 Lyon, France}
\address[UChile]{Departamento de F\'{\i}sica, FCFM, Universidad de Chile,
  Casilla 487-3, Santiago, Chile}

\begin{abstract}
We study, via hydrodynamic equations, the granular temperature profile of a
granular fluid under gravity and subjected to energy injection from a
base.  It is found that there exists a turn-up in the
granular temperature and that, far from the base, it increases linearly
with height. We show that this phenomenon, observed previously in experiments
and computer simulations, is a direct consequence of the heat flux
law, different form Fourier's, in granular fluids.  The positive granular temperature
gradient is proportional to gravity and a transport coefficient
$\mu_0$, relating the heat flux to the density gradients, that is
characteristic of granular systems. Our results provide a method to compute the
value $\mu_0$ for different restitution coefficients. The theoretical
predictions are verified by means of molecular dynamics simulations, and the value
of $\mu_0$ is computed for the two dimensional inelastic hard sphere model.
We provide, also, a boundary condition for the temperature field that is
consistent with the modified Fourier's law.
\end{abstract}

\end{frontmatter}

  In a granular system a small amount of energy is lost at each collision
between particles. This energy loss at the microscopic level leads
to macroscopic dynamics quite different from those of systems with
particles undergoing elastic collisions. 
  One of the surprising consequences of the dissipative dynamics is the
granular temperature inversion in fluidized granular systems under
gravity. The granular temperature ---the {\em temperature} from now on--- is defined
proportional to the mean kinetic energy per particle in the reference frame of
the fluid. It has been observed that an open system subjected to a permanent
energy injection from the base, exhibits a minimum of temperature at some
distance from the boundary, and from there on temperature increases with height.
In spite of the minimum the energy flux continues always pointing upwards.
This fact confirms the existence of a heat flux law, different from
Fourier's law, in these granular fluids. The heat flux is defined, for
granular fluids, as the kinetic energy flux in the reference frame of
the fluid. Nevertheless, the fact that the kinetic energy per
particle, that is the granular temperature, 
can increase with increasing height from the base is unexpected, given the
inelastic nature of collisions.

This temperature turn-up can be indirectly observed in experiments. For example,
in~\cite{Wildman} the authors report a temperature increase 30 times greater
than expected due to the slow convective movement they observe
in their experiments. Their conclusion is that this effect may not be
attributed to convection currents, and suggest that it could be
due to density gradients.
The temperature minimum has also been observed in computer
simulations~\cite{helal97,baldassarri00} and a recent article accounts
for this minimum with an hydrodynamic approach~\cite{Breynew}.

In this article, we deduce the presence of
the minimum. Instead of solving the whole
hydrodynamic equations for dilute granular gases as in~\cite{Breynew},
we show, using simple hydrodynamic arguments,
that the temperature minimum has its origin in the modification of the Fourier
law for the heat flux in granular
fluids~\cite{lun84,brey96a,brey98,sotomar99}. 
Also, unlike in~\cite{Breynew},  these simple hydrodynamic arguments
predict, far from the base, a linear increase of temperature with height
with a slope that can be deduced directly from the theory. This linear
dependence is verified using molecular dynamic simulations of a simple
granular model.

Consistently with these results, we establish the appropriate boundary
condition for the temperature of an unbounded system, and for a system
bounded by an upper adiabatic wall.
Finally we provide a method to measure experimentally the transport coefficient
associated  with the non-Fourier heat flux law.
 All these results are compared with molecular dynamics simulations for a
bi-dimensional system.

Hydrodynamic equations (that is, closed equations for a reduced number of
macroscopic fields), resembling those of elastic systems, have been derived for
the description of granular fluids~\cite{jenkins83,jenkins85c,brey96a}.
Even though the validity of such description might be controversial
\cite{kadanoff99,Goldhirsh}, hydrodynamic equations
together with appropriate constitutive relations have shown to be
valid in quasi-elastic low-density stationary states.

The difference between the elastic and the granular hydrodynamic equations
is an extra sink term in the energy balance equation which accounts for
the energy dissipation.  Also, constitutive relations are quite
different in granular materials and in elastic systems. Fourier's Law, as
mentioned, does not hold for these dissipative systems, but rather the heat
flux follows a law in the form~\cite{lun84,brey96a,brey98,sotomar99}

\begin{equation}
{\bf J} = -k \nabla T - \mu \nabla n, \label{heatlaw}
\end{equation}
${\bf J}$ being the heat flux, $k$ and $\mu$ the transport coefficients,
$T$ the granular temperature, and $n$ the number density. For small
dissipation and low density, kinetic theory predicts that $k$ and
$\mu$ must be of the form
\begin{eqnarray}
k   &=&k_0 \sqrt{T}, \nonumber    \\
\mu &=& \mu_0 \frac{T^{3/2}}{n}, \label{kmulown}
\end{eqnarray}
where $k_0$ and $\mu_0$ depend only on the dissipative coefficient and $\mu_0$
vanishes in the elastic limit. In quasi-elastic systems, $\mu_0\ll
k_0$~\cite{sotomar99}.

Experimentally, fluidized stationary states can be achieved
by means of a vibrating base with some amplitude and frequency. In
the low amplitude and high frequency limit, quasi-elastic systems reach
non-equilibrium stationary states in which the vibrating base plays the role
of a stationary boundary. A thermal wall is a useful theoretical model to
describe this kind of boundary: each time a particle collides with the wall, it
comes out with a velocity sorted out from a Maxwellian distribution.

While the results presented here should not depend on the particular mechanism
of the energy injection, for simplicity, we will focus on an open system
fluidized by a thermal base at the bottom.  The system is kept bounded by a
gravitational field pointing downwards with intensity $g$. 

Such a system may exhibit different regimes. If density, gravity
and dissipation are not too large, the system reaches a stationary state:
the conductive regime. This state is characterized by a vanishing velocity
field and density, temperature, and heat flux fields
depending only on the vertical coordinate $z$. For increasing 
density, dissipation and/or gravity the system starts developing different
instabilities such as Rayleigh Benard-like convection and
solidification~\cite{Rosa00}.

\begin{figure}[htb]
\begin{center}
\epsfig{file=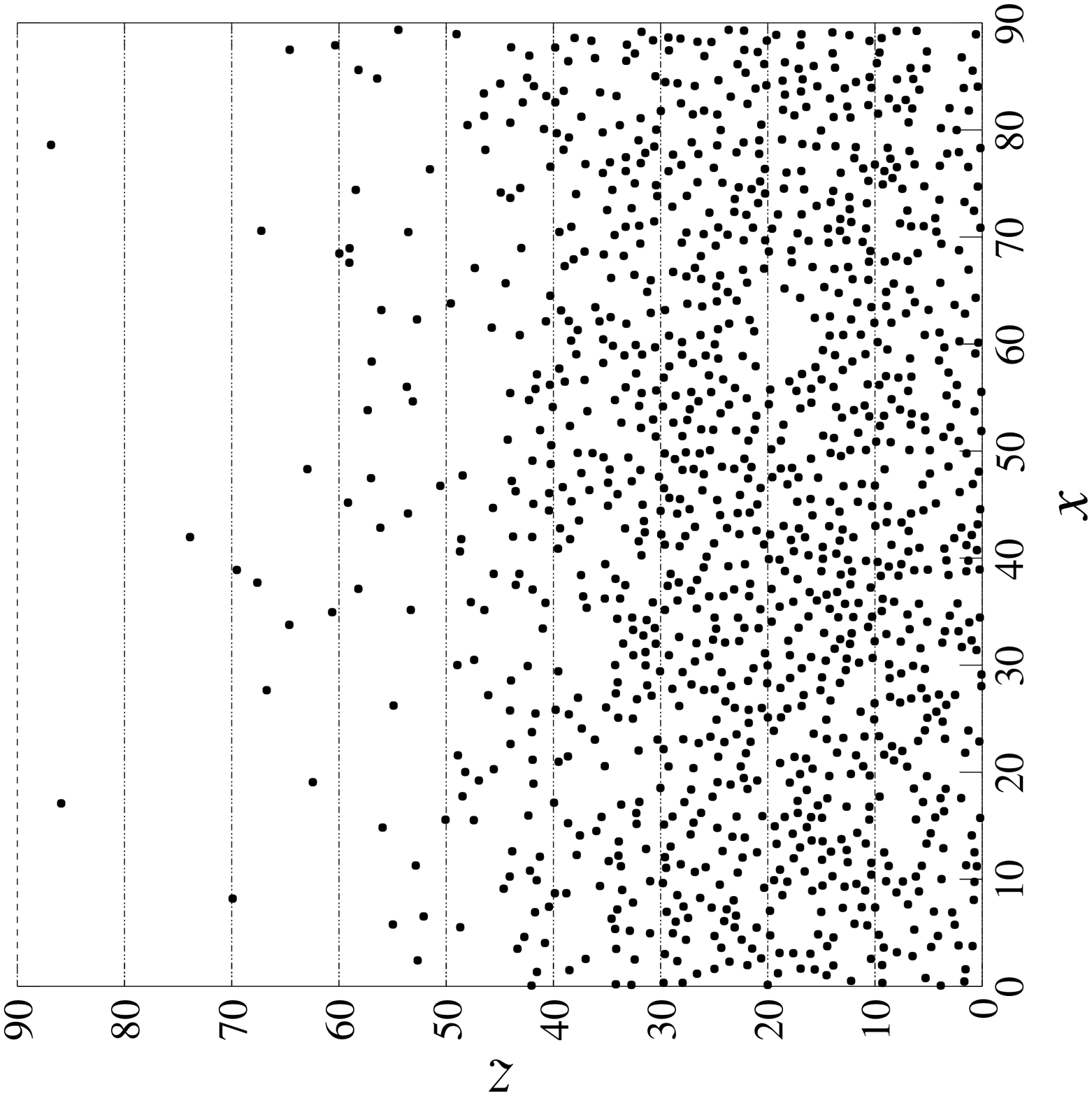,width=0.9\columnwidth,angle=270}
\end{center}
\caption{Snapshot of an open granular system obtained in a bi-dimensional
  molecular dynamics simulations of the inelastic hard sphere
  model. The system is fluidized by a thermal injection base and
  it is kept bounded by the gravity field pointing downwards.}
\label{Fig.Config}
\end{figure}

In the conductive regime, the energy dissipation at collisions induces a
vertical temperature gradient. Energy is injected through the base and
dissipated in the volume, with the corresponding decrease of the temperature
with height. A snapshot of a typical configuration in the conductive regime is shown in
Fig.~\ref{Fig.Config}.

Close to the base, we can state that the temperature decreases
with height due to the high dissipation rate. Depending on the gravity
acceleration and the inelasticity , the negative temperature
gradient, is sometimes accompanied by a density increase which,
at some extreme conditions, can lead to clustering and solidification.

Far from the base, on the other hand, we know that
the density decreases by the action of gravity.
Indeed, if the height is large enough ($z\ge 45$ in Fig.
\ref{Fig.Config}) we find just a few particles moving mainly in free flight,
so that we can safely consider that there is a height above which the
pressure is given by the ideal gas equation of state, $p=nT$.

The momentum balance equation states that the pressure obeys the barometric
law $\frac{dp}{dz} = -ng$, so that together with the ideal gas expression
for $p$ we obtain,
\begin{equation}
n \frac{dT}{dz} + T \frac{dn}{dz} = -ng . \label{bariometric}
\end{equation}

Inserting the expression for the density gradient derived
from~(\ref{bariometric}) into~(\ref{heatlaw}) and using the dilute limits
for the transport coefficients~(\ref{kmulown}), the heat flux far from the
base is given by
\begin{equation}
J = -\left(k_0 - \mu_0 \right) \sqrt{T} \frac{dT}{dz} + \mu_0 g
\sqrt{T} . \label{hetafluxg}
\end{equation}

As the density goes to zero with increasing height, the heat flux must
consequently vanish as there are basically no particles to transport energy.
It can be assumed then that there is a height where $J\ll \mu_0 g \sqrt{T}$.
Above this height the heat flux in Eq.~(\ref{hetafluxg}) can be neglected
leading to the following relation for the temperature gradient as $z$ goes
to infinity
\begin{equation}
\frac{dT}{dz} = \frac{\mu_0 g}{k_0-\mu_0} . \label{gradT}
\end{equation}

That is, the temperature gradient is a constant at high positions so that,
far from the heating base, we can predict that the temperature field
increases linearly with height with a slope proportional to $g$ and $\mu_0$.

Furthermore, as the temperature near the base decreases with height, it can
be deduced, by continuity, that the temperature field has a minimum. Above
the minimum, the temperature increases until it reaches the linear asymptotic
dependence given by Eq.~(\ref{gradT}).

\begin{figure}[htb]
\begin{center}
\epsfig{file=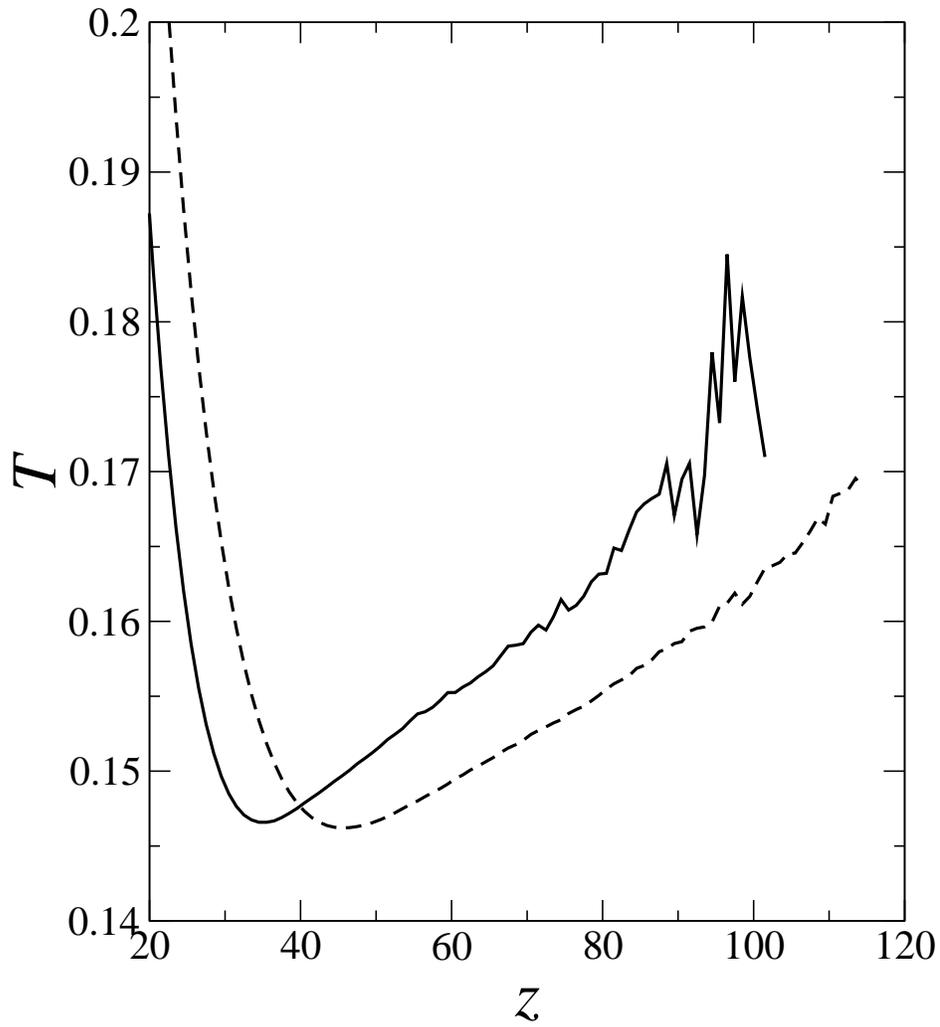,width=\columnwidth}
\end{center}
\caption{Temperature profile obtained from MD simulations. System A (dashed
curve) corresponds to $N=1120$, $L=89.44$, $q=0.01$, and $\hat{g}=0.02$.
System B (solid curve) corresponds to $N=560$, $L=44.72$, $q=0.01$, and
$\hat{g}=0.03$.}
\label{Fig.perfilT}
\end{figure}

This prediction is fully corroborated in our MD simulations, as observed in
Fig.~\ref{Fig.perfilT} where the measured temperature profile has been
plotted for two different systems. In both of them we clearly observe the
minimum and the further linear increase of the temperature.

In our MD simulations, the inelastic hard sphere model (IHS) in two
dimensions is used as the microscopic model for the particles.
 Grains are modeled as smooth hard disks that dissipate energy at each
collision through a constant dissipative coefficient $q$, related to the
normal restitution coefficient $r=1-2\,q$, so that $q=0$ represents the
elastic case. The simulated systems are composed of $N$ inelastic hard disks. In the
horizontal direction $x$, we set periodic boundary conditions.  The box
width is $L$. At the bottom we imposed a thermal wall at temperature $T_0$.
Due to the absence of an intrinsic energy scale in the IHS model, we fix the
temperature $T_0$ to unity. Results for other temperatures are obtained by
simple dimensional analysis. Also, units are chosen such that the disks
diameters $\sigma$ and masses $m$ are set to one. In these units the
dimensionless gravity is $\hat{g}=m\, g\, \sigma/\, T_0$. The system is
completely defined giving the values of $N$, $L$, $\hat{g}$, and the dissipative
coefficient $q$. For each value of the parameters we thermalize the system after
which we measure the macroscopic fields, averaging over a long simulation time.

In Fig.~\ref{Fig.perfilT} we present the results for: system A
(dashed line) with parameters $N=1120$, $L=89.44$, $q=0.01$, and $\hat{g}=0.02$;
and  system B (solid curve) with $N=560$, $L=44.72$, $q=0.01$, and
$\hat{g}=0.03$. The snapshot presented in Fig.~\ref{Fig.Config} corresponds to
the system A.  Comparing both figures it can be observed that the
temperature minimum (at $z\approx45\sigma$) lies close to the free surface
of the system but still into the bulk. The system size, under which 99\% of
the particles lie, is $60\sigma$ so that the observed linear dependence of
the temperature remains up to twice the system size, proving to be the real
asymptotic behavior. Similar behavior is observed in system B, that has an
effective height of $z=45\sigma$ and the temperature minimum is located at
$z=35\sigma$. For higher positions, the density is very low implying large statistical
errors.

 To test any possible finite size dependence on the temperature slope, we
performed simulations for different number of particles $N$, keeping fixed
the ratio $N/L$. We found that even if the profiles $n(z)$ and $T(z)$ may
suffer variations, the asymptotic temperature slope remains independent of
$N$ for $N\geq 280$.

 We could deduce from these results that the temperature could achieve
arbitrarily large values as height increases. However, our hydrodynamic
description is limited because of the zero density, or infinite Knudsen
number, limit at these heights.
Our prediction at high positions is, nevertheless, physically
meaningful as the combined effect of the temperature increase (linear) and the
density decrease (almost exponential) is a net decrease of the energy density
with height. So, even if the average kinetic energy per particle increases, the
total energy of the system remains always finite.
In summary, the prediction of the linear regime is valid up to some height after
which the starting equations (hydrodynamics) are not valid, but anyway give
physically meaningful predictions.

A system bounded by an upper adiabatic wall (that is, grains are reflected
elasticaly) is a similar, although not identical, problem to that of the
unbounded system. At an adiabatic boundary, the heat flux is exactly zero so
that Eq.~(\ref{gradT}) is the exact expression for the temperature gradient
at the wall if the density at the wall is low enough.  The temperature
gradient at the elastic wall is then positive, provided $\mu_0\ll k_0$. This
ensures the existence of a minimum, as the temperature gradient is negative
at the base. The minimum of temperature induced by an upper elastic wall has
been observed in one dimensional simulations ~\cite{Rosa01}.

It is clear that these results concern closely the
boundary condition for the temperature field that must be
used to solve the hydrodynamic equations in granular systems.
In an unbounded system under gravity, Eq.~(\ref{gradT})
must be imposed as the boundary condition for the temperature at infinity.
The same boundary condition applies to any adiabatic wall,
contrary to the usual condition $dT/dz=0$ that has traditionally
been used in computations of granular systems.

 Those experiments where the stationary temperature inversion has been
observed~\cite{Wildman}, have the normal lack of statistics at high
positions and the linear increase of temperature can not be easily detected,
specially if this is not the final purpose of the experiment.  We want to
stress that the measurement of the temperature slope, either experimentally
or in simulations, provides a method to obtain the transport coefficient
$\mu_0$ for small dissipation coefficients.

As an example, we have calculated the value of $\mu_0$, at the lowest order
in the dissipative coefficient $q$, for the two dimensional inelastic hard
sphere model.  We consider systems with $N=560$ and $L=44.72$. In order to
avoid instabilities --especially the convection described in
Ref.~\cite{Rosa00}-- systems with low dissipative coefficients and low
gravity values have to be considered. Our simulations have been done with
$\hat{g}=0.01$ and $\hat{g}=0.02$, and $0.002\leq q \leq 0.02$.

For these small values of $q$, we can use $(k_0-\mu_0)\approx k_0$ to first
order in $q$ . Also, in the same order of approximation, $k_0$ can be set
to that of the elastic hard disk system, $k_0=2/\sqrt{\pi}$, neglecting the
linear correction in $q$. Then, $\mu_0$ is given by
\begin{equation}
\mu_0 = \frac{2}{\sqrt{\pi}\,g}\frac{dT}{dy}, 
\end{equation}
expression that is valid only up to order $q$.

\begin{figure}[htb]
\begin{center}
\epsfig{file=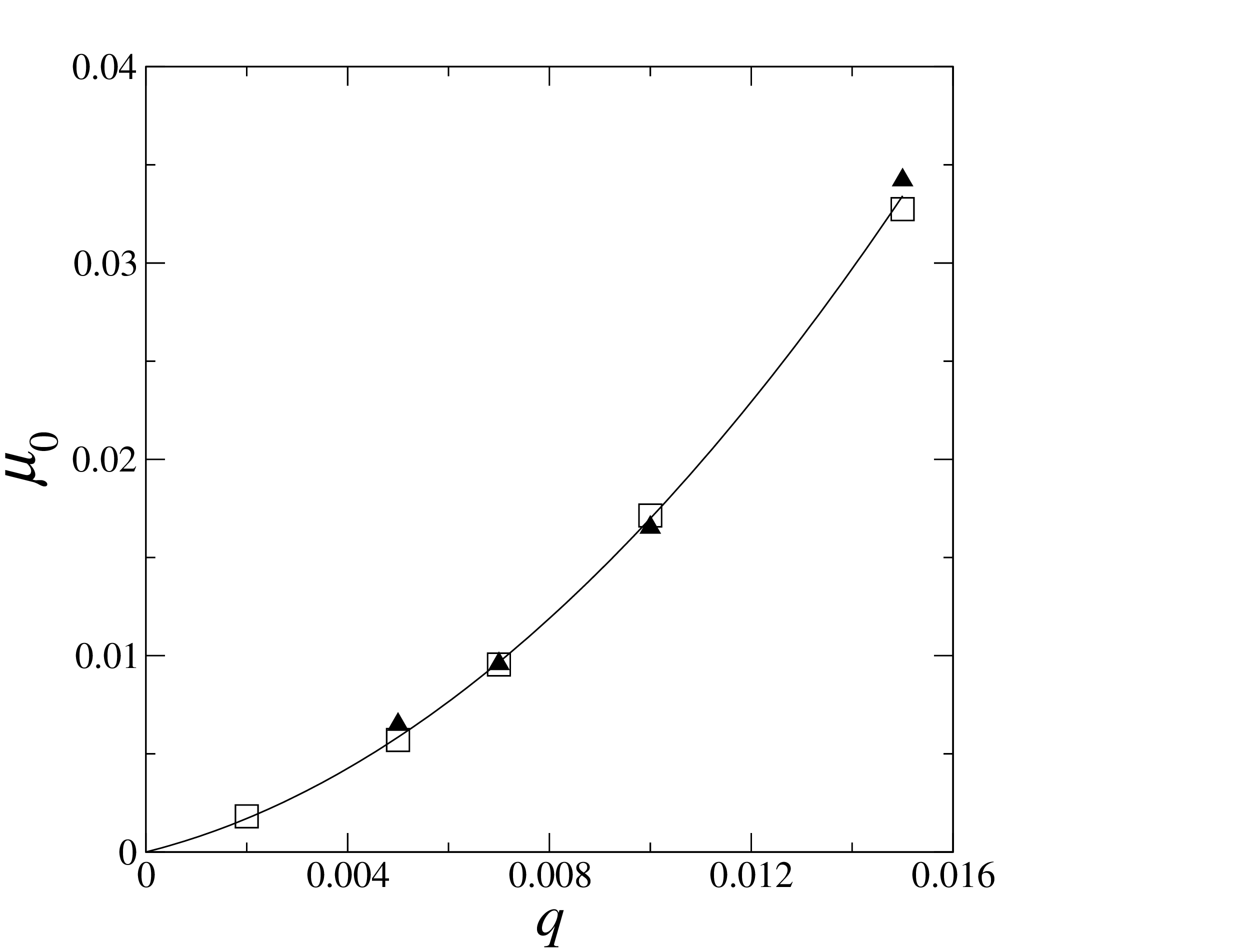,width=0.75\columnwidth,angle=270}
\end{center}
\caption{Computed values of $\mu_0$ as a function of $q$ obtained in
  two series of simulations. Solid triangles correspond to $\hat{g}=0.01$ and
  open squares to $\hat{g}=0.02$. The solid curve is a quadratic fit
  in $q$ of the
  obtained data.}
\label{Fig.muq}

\end{figure}

In Fig.~\ref{Fig.muq}, we present results for $\mu_0$ obtained in the
simulations. It can be observed that the two sets, with different values of
$\hat{g}$, collapse into the same curve, confirming the linearity with 
$\hat{g}$ of the temperature  gradient.
  
As expected, $\mu_0$ goes to zero in the elastic limit. The computed values
of the two series are fitted to a single quadratic function in $q$ where it
is imposed that $\mu_0$ must vanish for $q=0$. The obtained fit is 
\begin{equation}
\mu_0 = 0.64 q + 106 q^2 \label{munumeric2}
\end{equation} 
noting that only the linear term in the last fit is
consistent with the approximation done.
In Ref.~\cite{sotomar99} the 
authors computed $\mu_0$ using a different method from that described in here,
with the result $ 
\mu_0 = 0.7 q \label{munumeric} $
which agrees with the result found in the present paper. Nevertheless,
we want to point out that the present method for computing $\mu_0$ uses the 
temperature profile over a wide range (the region where $T$ increases 
linearly) instead of using the information over just one point as in 
Ref.~\cite{sotomar99}. Therefore this method of computing $\mu_0$ is more 
accurate in the dilute case.

In the highly fluidized regime, the particular energy injection mechanism
becomes irrelevant far from the energetic boundary. This implies that even
in a system on a vibrating base, the method presented here should be fully
valid. Also, even though our predictions are valid in the dilute case, we expect
that in denser regimes there should be also a temperature turn-up and a
temperature increase with height (though maybe not linear). In this case, there
are explicit density dependences in the equation of state and transport
coefficients, giving rise to a more complex differential equation for $T$
compared to~(\ref{gradT}). Nevertheless, as long as there is a non-vanishing
coefficient $\mu$, the equation is not trivial and a temperature gradient
is predicted.

In summary, we have shown that the granular temperature turn-up observed in
vibrofluidized granular systems has its origin in the modification of the
Fourier law for the granular heat flux. 
The vanishing energy flux far from the base together with the density
gradient produced by the gravity  imposes a positive and constant
temperature gradient. Careful measurements of this
gradient provide a method to compute the transport coefficient $\mu_0$,
either experimentally or in simulations.

R.S. acknowledges the grant from {\em MIDEPLAN}, Chile, the {\em
Programa de inserci\'{o}n de cient\'{\i}ficos chilenos} of 
Fundaci\'{o}n Andes, and the support of the
FONDECYT project 1010416. R.R. acknowledges
{\em Marie Curie Individual Fellowships} from the European Community.

\newpage

\begin{thebibliography}{1}
\bibitem{helal97}
K. Helal, T. Biben, and J.~P. Hansen, Physica A {\bf 240},  361  (1997).
In this article, the authors present an explanation for the temperature
turn-up that implies a heat flux coming from infinity, which is clearly
inconsistent with the absence of an upper wall, as it was the case.
\end{thebibliography}
\end{document}